# A dialectical view of the variational principles of mechanics

Qiuping A. Wang

Institut Supérieur des Matériaux et Mécaniques Avancés du Mans,

44 Av. Bartholdi, 72000 Le Mans, France

Abstract

The aim of this work is to provide a possible philosophical motivation to the variational principles of physics in general and a possible way to unify the axiomatizations of mechanics theories. The leitmotif of this work is a dialectical view of the world: any smooth motion in nature is the consequence of the interplay and dynamical balance between pairs of opposite elements of the motion. We stress that the opposite property here differs from the common sense of opposing elements (of conflicting appearance but of same nature) such action/reaction forces between two bodies. The opposite elements here are incompatible and mutual exclusive in nature, with contrasting and complementary characteristics such as active/passive, disorder/order, curvature/flatness, etc. The dynamical balance is defined by an invariant variational relationship between all the pairs of opposite elements active in the considered motion. From this idea, an axiomatization of quantum mechanics is proposed in addition to the derivation of several existing variational principles of mechanics.

## 1) Introduction

The variational approach in physics, such as the principles of least action [1][2] and of virtual work[3][4] for mechanics as well as the principle of maximum entropy [5][6] for thermodynamics, is widely used for the axiomatic formulation of physical theories. It is also, with a long historical development accompanied by philosophical and mathematical controversies, a philosophy laden subject whose last word is not yet spoken to date.

Variational principles have a strong metaphysical origin[1][2] and were developed under the influence of teleological tendency of the pioneers (Fermat, Leibnitz, Maupertuis, Euler and others)[2][7][8] who used minimum calculus to express the longstanding belief of nature's (or God's) perfection. The essential of their approach was to create virtual world of possible events and then to select only a small number of these by a mathematical extremum believed to express a *purpose*[1][2][7][8]. For this view point, nature is a selective system similar to a biological system. This teleological view was rejected gradually in nineteen and twentieth centuries by the anti-metaphysical trends of empiricism and related philosophical circles[7][8], for which the variational principles are equivalent to differential equations and have only esthetic and economic interest in the adaptation of thoughts to "elementary facts"[7][8], i.e., in axiomatic formulation of physics theory.

Since this moment, variational principles have gradually lost its metaphysical bedrock and become completely devoid of philosophical and physical motivation. The old questions, answered in the past with the belief of God's intelligence, have been left unanswered. Some of these fundamental inquiries are: why the maximum/minimum is a criterion of nature for selecting events; why the time integral of Lagrangian (the difference between kinetic and potential energy) has extremum for real motion; why entropy is maximal for stable probability distribution[1], and why entropy production should be maximal or minimal for nonequilibrium process. To these questions and many others about the origin of the maximum/minimum calculus, we may add a final one: whether or not there is a common cause to so many seemingly different, sometimes mutually exclusive principles in nature? This mutual exclusivity has a very illustrative example: the conflict between the least action principle and the maximum entropy (and its production) principle: the former yields motions satisfying

[1] This principle has been more or less justified with non mathematical arguments by Jaynes[5] and Tribus[9] who considered probability in the Bayesian sense as a subjective quantity depending on our incomplete knowledge about given situation. This interpretation of probability encountered enormous resistance from the physicists believing the objective frequency interpretation of probability, for which the principle remains a mysterious hypothesis.

Liouville's theorem and Poincaré's recurrence theorem, the second uses the notion of entropy, a quantity whose evolution law (second law of thermodynamics) is believed to be at odds with the above mentioned theorems[10].

Since the rejection of teleological view, physicists and philosophers have tried to answer these above questions. For many, the superiority of variational principles with respect to differential equations is obvious[8], but they failed to grasp the matters that would be other than a formal superiority in mathematical description and capable of deepening our understanding of nature. As far as we know, no conclusive statement can be made to date about the very reality behind the variation principles in physics.

Efforts have been made to derive variational principles of physics from other general philosophical hypothesis. For example, the variational principles was attributed to the principle of unique determinacy[7] with the slogan "nature happens only ones"[11]. If this is true in the case of regular motion determined uniquely by maximum/minimum calculus, it is not the case for irregular (random or chaotic) dynamics, since there may be infinite number of possible events taking place at a given moment, or infinite number of different paths (observable with non zero probability) for a system starting from a sole state (this randomness of the motion should be quantitatively tackled in the statistical calculus of variation, see section 5 below). We think it is worthy to mention here the effort of Feynman to derive least action principle from quantum mechanics within his path integral formalism with the factor $e^{iA/\hbar}$[12]. The argument is roughly that, in the propagator from one point to another calculated by path integral, the main contribution of the factor $e^{iA/\hbar}$ is only around the least action path when the action $A$ is much larger than Planck constant $\hbar$. This reasoning is plausible. The only difficulty is that the origin of the factor $e^{iA/\hbar}$ for quantum system, inspired by an observation of Dirac about quantum phase[13], is still a mystery to date.

The present work is a tentative to fill the void of understanding variational principle by introducing a dialectical view of the duality of nature. This view is: any entity or events contains pairs of opposite, mutually exclusive in nature but nevertheless equally essential and complementary parts. This very ancient and long-standing philosophical point of view is not only a conceptual conviction, but also a belief, common to both occidental and oriental cultures, stemming directly from observation of natural and social movements.

In its modern development accomplished mainly by Hegel and Marx, the dialectic often stresses conflict, negation and change of the opposite parts. However, one of the inevitable and essential aspects of any motion is the smooth and continuous evolution state with dynamical balance between opposites. This is also an important aspect of dialectic. In fact, in this work, we will concentrate on the smooth and continuous motion of physical system using this notion of harmonic balance to develop a variational calculus. By invariance consideration, this balance is mathematically defined by a variational relationship between the opposite quantities. The main idea is: any physical phenomenon in *smooth* motion is the consequence of the *interplay* and *invariant balance* of pairs of opposite elements. To our opinion, this may be the very reality hidden behind the different variational principles in physics. Examples of derivation of several variational principles are given. It is hoped such approach can help us to understand more about why things happen that way in physics.

## 2) Dialectical variational principle

In this section, we will develop a dialectical variational principle (DVP) on the basis of pairs of opposite elements classified into two categories of quantities having completely different and opposing natures and behaviors. One category has the passive, ordered, hidden or straight (flat) characters; the other one has completely opposite characters such as active, disordered, open or curved character. In what follows, we will use this attribute as an essential criterion for the choice of opposites.

The DVP is stated as follows:

> *Any event of a system never happens without the occurrence of the invariant balance between all pairs of opposite, mutually exclusive in nature but complementary, mutually rooted and transformable parts of the system that undergo variations due to the event.*

Let us spell out the idea by giving the following definitions.

a) *Yi* is defined as a physical quantity charactering the passive, ordered, inanimate, hidden or flat aspects of a motion.

b) *Ya* is defined as a physical quantity charactering the active, disordered, animate, open or curve aspects of the same motion.

It should be stressed here that *Yi* and *Ya* must be empirically measurable quantities or mathematical expressions which can be univocally related to measurable quantities. Conceptual notions or abstract words devoid of physical instance (such as *black* and *white*, *up* and *down*, *large* and *small*, etc) are not considered. As defined below, *Yi* and *Ya* of mechanical system may be associated with, respectively, spatial accumulation of inertial force and driving force (work), time accumulation of potential energy and kinetic energy, or statistical expectation of random variables (representing order and observability) and entropy (measuring disorder and randomness), among many others. Examples using these pairs of opposite quantities will be given below.

In general, for given situation, these opposing characters are absolute, meaning that *Yi* and *Ya* cannot be defined from physical quantities of same nature. However, in different situation, a given quantity can be either related to *Yi* or *Ya* depending on the events containing it. Inertial force, for example, is rather *Yi* in accelerating motion but *Ya* in decelerating dissipative motion since it maintains the motion, an active aspect of the process.

c) The variation $\delta Y$ of an element *Y* is a virtual change without time passage and confined by constraints.

One of the advantages of the use of variation $\delta Y$ of a quantity *Y* is that $\delta Y$ can be chosen in independent way from the frame of reference, while this is not possible a priori for the magnitude *Y* itself which is in general dependent on its frame. The use of variation makes it possible to establish a universal balance or equilibrium independent of the frame of reference as follows.

d) Variational Balance: for a virtual process yielding variation of the opposite elements $Yi_l$ and $Ya_l$, the invariant balance between the pairs of opposites is described by

$$\sum_l \delta Yi_l - \sum_l \delta(\lambda_l Ya_l) = 0 \qquad (1)$$

where $\lambda_l$ for a given pair *l* is a coefficient characterizing the interplay and the equilibration of the pair *l*. The value of $\lambda_l$ should depend on the nature of the pair *l* as well as on the property expressed in this equation (balance, complementarity, transformability, etc.). The reader will see that this parameter can be determined in each application of the principle.

If necessary, constraints $C_k$ on the balance may be added à la Lagrange such that:

$$\sum_l \delta Yi_l - \sum_l \delta(\lambda_l Ya_l) + \sum_k \delta(\alpha_k C_k) = 0 \qquad (2)$$

where $\alpha_k$ is a multiplier related to the constraint $C_k$. However, Eq.(1) can be used even when there are constraints, meaning that the constraints can be taken into account in the subsequent treatments.

By using these above definitions, DVP can be stated as follows:

> *Any continuous and smooth event of a system never happens without the occurrence of variational balance between all pairs of Yi and Ya that undergo variation due to the event.*

The application of DVP consists, first of all, in the identification of pairs of opposite elements whose interplay determines the event, as shown below.

## 3) Principle of least action

We consider a system with Hamiltonian $H(q,\dot{q},t) = T(\dot{q},t) + V(q,t)$ where $T$ is the kinetic energy depending on time and velocity, and $V$ the potential energy depending on the configuration of the system as well as time. We suppose, without loss of generality, that in a given frame of reference, the system is in non uniform motion with acceleration $\vec{a}_i$ of the point $i$. $m_i$ is the mass of the point having the force $\boldsymbol{F}_i$ ($i$=1,2,…$N$). Now we have to identify the pairs of opposite elements to put into Eq.(1). Let $V_i$ be the potential energy associated with $\boldsymbol{F}_i$. $V_i$ represents, at a given moment of time and in a given frame of reference, an immobile and hidden motion (or latent capacity of doing work, *vis mortua*). Let $T_i$ be the kinetic energy which represents a mobile energy and apparent motion (evident capacity of doing work and of moving against resistance, *Vis viva*). Hence it is evident that $V = \sum_{i=1}^{N} V_i$ (total potential energy) and $T = \sum_{i=1}^{N} T_i$ (total kinetic energy) are two essential opposite elements of the motion at any given moment of time which satisfy the definition of *Yi* and *Ya*. However, $V$ and $T$ will not be considered as *Yi* and *Ya*. We choose instead the time accumulation of them $\int_{t_a}^{t_b} V dt$ and $\int_{t_a}^{t_b} T dt$ for the case of a motion between a point $a$ and point $b$ in phase space during a given period of time $\tau = t_b - t_a$. This choice has two reasons. The first is the motion during $\tau = t_b - t_a$ should be taken into account as a whole. That is, if we want to see the effect of a virtual deformation on the motion, we should calculate the entire effect over the whole motion, not only the instantaneous effect over a temporary state characterized by the instantaneous energy. This first reason coincides with the second one, that is, according to Courant and Hilbert, the object of the calculus of variation "is to find extrema of functionals rather than extrema of functions of a finite number of independent variables. By a functional we mean a quantity or function which depends on entire course of one or more functions" [14].

Now let us define

$$Yi = \int_{t_a}^{t_b} Vdt \text{ and } Ya = \int_{t_a}^{t_b} Tdt . \tag{3}$$

Eq.(1) then reads

$$\delta \int_{t_a}^{t_b} Vdt - \lambda \delta \int_{t_a}^{t_b} Tdt = 0 \tag{4}$$

which is nothing but the usual least action principle (we must let $\lambda = 1$):

$$\delta \int_{t_a}^{t_b} (T - V)dt = \delta A = 0 \tag{5}$$

with the Lagrange action $A = \int_{t_a}^{t_b} Ldt$ where $L = T - V$ is the Lagrangian. As well known, this principle yields the Lagrange equation for analytical mechanics :

$$\frac{\partial L}{\partial x_i} - \frac{\partial}{\partial t}\left( \frac{\partial L}{\partial \dot{x}_i} \right) = 0 . \tag{6}$$

(with only one degree of freedom $x_i$ of the point *i*). As well known, the principle in Eq.(5) has been used for the axiomatic formulation of almost all the physics theories including relativity theory, electromagnetism theory, geometrical and wave optics. However, the situation concerning quantum mechanics is somewhat ambiguous since in Feynman's formalism, the least action principle does exist but only in the limit $\hbar \to 0$ [12]. The formulation based on the factor $e^{iA/\hbar}$ cannot be directly derived, as far as we know, from variational calculus using Lagrange action.

## 4) Principle of virtual work

That the above analysis is based on the cumulative energy of the motion during a period is natural since a motion should be in principle tackled according to its global effect. As a matter fact, this global treatment finally yields Eq.(6) which is an instantaneous description of the motion with differential equation. From this observation, basic principles regarding only instantaneous effect should be possible. One of these is the so called Lagrange-d'Alembert principle of virtual work. To see this, we write Eq.(5) as

$$\delta \int_{t_a}^{t_b} L dt = \int_{t_a}^{t_b} \sum_{i=1}^{N} \left( \frac{\partial L}{\partial x_i} - \frac{\partial}{\partial t} \left( \frac{\partial L}{\partial \dot{x}_i} \right) \right) \cdot \delta \vec{r}_i dt = 0 . \tag{7}$$

where $\delta \vec{r}_i$ is the virtual displacement of a point *i*. Let us introduce a virtual velocity $\vec{v}_i = \frac{\delta \vec{r}_i}{\delta \tau_i}$, the above equation reads

$$\int_{t_a}^{t_b} \sum_{i=1}^{N} \left( \frac{\partial L}{\partial x_i} - \frac{\partial}{\partial t} \left( \frac{\partial L}{\partial \dot{x}_i} \right) \right) \cdot \vec{v}_i \delta \tau_i dt = 0 . \tag{8}$$

where $\delta \tau_i \big|_a = \delta \tau_i \big|_b = 0$. According to the calculus of variation[8], it follows that

$$\sum_{i=1}^{N} \left( \frac{\partial L}{\partial x_i} - \frac{\partial}{\partial t} \left( \frac{\partial L}{\partial \dot{x}_i} \right) \right) \cdot \vec{v}_i = \sum_{i=1}^{N} (\vec{F}_i - m a_i) \cdot \vec{v}_i = 0 . \tag{9}$$

This is the principle of virtual velocity equivalent to the Lagrange-d'Alembert principle of virtual work[4]

$$\delta W = \sum_{i=1}^{N} (\vec{F}_i - m_i \vec{a}_i) \cdot \delta \vec{r}_i = 0 . \tag{10}$$

As a matter of fact, this equation can be derived directly from Eq.(7) if we consider only the motion in the infinitesimal time internal $t \to t + dt$.

When the motion is uniform without acceleration: $\vec{a}_i = 0$, Eq.(10) reads

$$\delta W = \sum_{i=1}^{N} \vec{F}_i \cdot \delta \vec{r}_i = 0 \tag{11}$$

which characterizes static equilibrium. We can extend the term *equilibrium* characteristic of Eq.(11) to the dynamical state characterized by Eq.(10) and refer to it as *dynamical equilibrium*, a term which will be useful for the thermodynamic equilibrium discussed below.

On the other hand, Eq.(10) can be obtained directly from DVP if we consider *Yi* as the work (spatial accumulation) of the initial forces, i.e., $Yi = \sum_{i=1}^{N} \int_a^b m_i \vec{a}_i \cdot d\vec{s}_i$ and *Ya* that of the driving forces $Ya = \sum_{i=1}^{N} \int_a^b \vec{F}_i \cdot d\vec{s}_i$ from point *a* to point *b*, where $ds_i$ is the elementary displacement of the point *i*. Then the variation of *Yi* due to the virtual displacements $\delta\vec{r}_i(t)$ at a given moment *t* must be

$$\delta Yi = \sum_{i=1}^{N} m_i \vec{a}_i(t)\cdot\delta\vec{r}_i(t) \tag{12}$$

and similarly

$$\delta Ya = \sum_{i=1}^{N} F_i(t)\cdot\delta\vec{r}_i(t)\,. \tag{13}$$

Hence DVP directly yields Eq.(10) with $\lambda = 1$.

## 5) Maximum path entropy principle for nonequilibrium thermodynamics

We still consider non dissipative system. The system is now perturbed by noises and has a random (irregular) dynamics. An essential difference between this irregular dynamics and the regular one is that the path from a point *a* to a point *b* for a given period of time is no more unique, as discussed in our previous work[15]. At any moment of time, a system in irregular dynamics does not have only one state, contrary to the case of regular dynamics. In the case of thermodynamic system containing a large *N* particles, the different states at any given moment are called microstates (*j*=1,2, …) each having a likelihood to be visited by the system. In this case, in addition to the pairs of opposite parts related to regular forces and kinetic energy, the aspects such as disorder and order, represented by some measurable quantities, of the motion are expected to play a role in the variational balance.

This role of order and disorder in the variational calculus can be shown as follows. Let $p_j$ be the probability of occurrence of *j* microstate at a given moment of time. The expectation of a quantity *Y* at this moment is given by $\overline{Y} = \sum_j p_j Y_j$ where $Y_j$ is the value of *Y* at the state *j*. Suppose now a variation $\delta Y_j$ of $Y_j$ in the state *j*. The regular or observable quantity of *Y* is the expectation $\overline{Y}$. It is obvious that the variation $\delta\overline{Y}$ is not only determined by $\delta Y_j$ since $\delta\overline{Y} = \sum_j p_j \delta Y_j + \sum_j Y_j \delta p_j$. So between the regular variation $\delta\overline{Y}$ and the expectation of microstate variation $\overline{\delta Y} = \sum_j p_j \delta Y_j$, there is the variation of a quantity $\Omega$ given by

$$\delta\Omega = \sum_j Y_j \delta p_j = \delta\overline{Y} - \overline{\delta Y} \tag{14}$$

It has been proven[16] that this quantity $\Omega$ is a measure of uncertainty associated with a random variable $Y$ and referred to as 'varentropy' in order to distinguish it from the usual information or entropy defined with given formula. It is straightforward to show that[17], in the case of equilibrium thermodynamic system having internal energy as random variable, $\delta\Omega$ is the heat transfer directly related to the thermodynamic entropy of the second law. This example shows well the role of disorder (irregularity represented by $\Omega$) and order (regularity and observability represented by the statistical expectation $\overline{Y}$).

Now let us extend DVP to the irregular dynamics in the case of a motion between two given points (microstates) *a* and *b* in phase space. Suppose a large number of possible paths *h* (*h*=1,2, …) for a given period of motion, each having a probability $p_h(a,b)$ of occurrence. Due to the perturbation of noise, the variational balance over each path is in general lost. So we do not have Eq.(4) for every path. As stated in the principle, the essential element for the balance is the variation but not the magnitude of the opposite quantities, hence we calculate the expectation of the variations $\delta Yi_h$ and $\delta Ya_h$ over all the paths such that

$$\overline{\delta Yi} = \sum_h p_h(a,b)\int_{t_a}^{t_b} \delta V_h dt \text{ and } \overline{\delta Ya} = \sum_h p_h(a,b)\int_{t_a}^{t_b} \delta T_h dt . \tag{15}$$

Then the application of DVP to these expectation $\overline{\delta Yi} - \lambda\overline{\delta Ya} = 0$ yields ($\lambda = 1$)

$$\sum_h p_h(a,b)\int_{t_a}^{t_b} \delta V_h dt - \sum_h p_h(a,b)\int_{t_a}^{t_b} \delta T_h dt = 0 . \tag{16}$$

This can be straightforwardly written as $\sum_h p_h(a,b)\delta A_h = 0$ or

$$\overline{\delta A_{ab}} = 0 . \tag{17}$$

Here $A_h(a,b) = \int_{t_a}^{t_b} L_h dt$ and $L_h = T_h - V_h$ is respectively the action and the Lagrangian of the system on the path *h*. Eq.(17) is the stochastic least action principle proposed in [15].

As shown in our previous work, Eq.(17) is equivalent to another expression:

$$\delta S_{ab} - \eta\delta\overline{A}_{ab} = 0 \tag{18}$$

where $\overline{A}_{ab} = \sum_h p_h(a,b)A_h$ is the expectation of action over all possible paths and $S_{ab}$ is the uncertainty in the choice of paths by the system and referred to as path entropy between *a* and *b*. It is defined by

$$\delta S_{ab} = \eta(\delta\overline{A}_{ab} - \overline{\delta A}_{ab}) = \sum_h A_h \delta p_h(a,b) . \qquad (19)$$

Hence, obviously Eq.(18) is an expression equivalent to Eq.(17) for random motion. $\eta$ is a characteristic constant of the process. For example, in the case of a Brownian motion, $\eta$ is inversely proportional to the diffusion constant[15].

As a matter of fact, $S_{ab}$ and $\overline{A}_{ab}$ can be considered as another pair of opposite elements of the motion and represent, respectively, the disorder (uncertainty, randomness) and order (certainty, observability). From this consideration, Eq.(18) is natural if we apply DVP to $Yi = \overline{A}_{ab}$ and $Ya=S_a$ and $\lambda = \eta$ .

Since the variation of path entropy is due to the variation of action, the functional property of the entropy can be discussed with respect to the action. The concavity of $S_{ab}$ was proved in [15]. So Eq.(18) is a maximum path entropy algorithm. This maximum calculus was an assumption in our previous work. But here it is a consequence of DVP or of the stochastic least action principle Eq.(17).

Note that the functional form of path entropy is not necessary in this principle for the maximum to be valid. This generic maximum varentropy algorithm (maxvent) makes it possible to maximize any appropriate varentropy functional (not only the Shannon one) that have physical meaning. The reader can find several examples of these we found directly from observed probability distributions[16]. If the entropy functional is given, maxvent calculus will yield corresponding probability distribution. Appendix I shows an example with Shannon varentropy yielding exponential path distributions.

## 6) Maximum entropy principle for equilibrium thermodynamics

In this section, we are interested in the variational calculus relative to the instantaneous dynamical uncertainty associated with equilibrium probability distribution. The measure of this uncertainty is still defined by Eq.(14) with one or more random variables [17] of equilibrium states. According to the above application of DVP to disorder and order, we can already write a variational algorithm such as

$$\delta\Omega - \sum_m \lambda_m \delta\overline{X}_m = 0 \qquad (20)$$

where $m$ is the index of the involved random variables and $\delta\Omega = \sum_m \left(\delta\overline{X}_m - \overline{\delta X}_m\right)$. $X_m$ can be determined or inferred through the nature of the system. For example, for a canonical ensemble in equilibrium state with $w$ microstates $j$ (1,2, … $w$) each having the probability $p_j$, the sole $X_1$ ($m$=1) should be the system energy $E_j$ according to the state principle[19].

Eq.(20) can be of course derived from the energy definition of $Yi$ and $Ya$. In what follows, we consider the $\delta Yi$ and $\delta Ya$ given by Eqs.(12) and (13). Using the same mathematics as in Eq.(15) with the probability $p_j$ of microstates, we obtain the expectation of $(\delta Yi)_j$ and $(\delta Ya)_j$ depending on the microstates $j$ of the system :

$$\overline{\delta Yi} = \sum_j p_j \left( \sum_{i=1}^{N} m_i \vec{a}_i \cdot \delta\vec{r}_i \right)_j \text{ and } \overline{\delta Ya} = \sum_j p_j \left( \sum_{i=1}^{N} \vec{F}_i \cdot \delta\vec{r}_i \right)_j . \qquad (21)$$

Then the application of DVP to these expectation $\overline{\delta Yi} - \lambda\overline{\delta Ya} = 0$ yields:

$$\sum_j p_j \left( \sum_{i=1}^{N} (\vec{F}_i - m_i \vec{a}_i) \cdot \delta\vec{r}_i \right)_j = 0 \quad \text{with } (\lambda = 1) \qquad (22)$$

By using this equation, a rather tedious calculation has been carried out in reference [17] for equilibrium thermodynamic of Grand-canonical, canonical and microcanonical ensembles. In what follows, we only briefly review the result for canonical ensemble as illustration.

It is straightforward to calculate that $(\vec{F}_i - m_i \vec{a}_i) \cdot \delta\vec{r}_i = -\delta e_i$ where $e_i$ is the energy of the particle $i$. Eq.(22) reads $-\sum_j p_j \left( \sum_{i=1}^{N} \delta e_i \right)_j = -\sum_j p_j \delta E_j = -\overline{\delta E} = 0$ for canonical ensemble. By the consideration of the first and second laws of thermodynamics of equilibrium system, it was proved that the virtual work principle, i.e., $\overline{\delta E} = 0$ led to

$$\delta(S - \beta\overline{E}) = 0 \qquad (23)$$

where $S$ is the thermodynamic entropy defined by $\delta S / \beta = \delta\overline{E} - \overline{\delta E}$ and $\beta$ is the inverse temperature[17]. This is the principle of maximum entropy (maxent) for thermostatistics. When derived in this way, the definition of the entropy functional is not necessary for the maxent. It may take if necessary other forms[16] different from the Gibbs-Shannon formula used in the formulation of statistical mechanics by Jaynes[5] and Tribus[9]. In their formalism, statistical mechanics is an inference theory instead of a theory of physics, and the most fundamental gradient of the formalism, maxent, is only an inductive method but not a deductive law of physics. In our formulation, maxent, and more generally speaking, maxvent, are no doubt laws of physics due to their tight kinship with the most fundamental law of mechanics such as virtual work principle and least action principle.

## 7) An axiomatization of quantum mechanics with variational principle

The axiomatic formulation of quantum mechanics has always been a dream of many physicists. No definitive result is obtained to date. Additional hypothesis are always used in the derivation of the Schrödinger equation. Here we would like to review the point of view considering Schrödinger equation as diffusion equation with complex diffusion constant and then to suggest a variational formulation of quantum mechanics. Complex diffusion is a well known effort which can be traced back to the initiative of Fürth in 1933[20]. The complex diffusion constant, just as the imaginary time and imaginary energy in certain formulation[21], is not a trivial hypothesis. We cannot explain why a diffusion may be complex. But if it is taken for granted, it would be possible to formulate quantum mechanics on the basis of the least action principle of Eq.(17), with $\eta = -\frac{i}{\hbar}$ and $i = \sqrt{-1}$. This, by Eq.(24) of appendix I, yields $\tilde{p}_h(a,b) = \frac{1}{Z_{ab}} e^{\frac{i}{\hbar} A_h(a,b)}$, the Feynman factor of path integral. However, this formulation has several nontrivial consequences about probability and entropy. The first one is that the transition path probability $\tilde{p}_h(a,b)$ from $a$ to $b$ in phase space is complex. We can nevertheless normalize it formally by path integral $\sum_h \tilde{p}_h(a,b) = 1$ over all possible paths from $a$ to $b$. The second consequence of complex diffusion is that the path entropy measuring the uncertainty in $\tilde{p}_h(a,b)$ is a complex information, hence Eq.(25) becomes $\tilde{S}_{ab} = \ln Z_{ab} - \frac{i}{\hbar}\overline{A}_{ab}$ where $Z_{ab} = \sum_h e^{\frac{i}{\hbar} A_h(a,b)}$ and the expectation of action $\overline{A}_{ab} = \frac{1}{Z_{ab}} \sum_h A_h(a,b) e^{\frac{i}{\hbar} A_h(a,b)}$ are in general complex numbers.

The physical attributes of probability having complex values is still an open question. But all the same $\tilde{p}_h(a,b)$ gives an estimation of the likelihood of the path $h$ from $a$ to $b$, taken into account in path integral. In this sense, the generic varentropy $\tilde{S}_{ab}$ defined by $\delta\widetilde{S}_{ab} = -\frac{i}{\hbar}(\delta\overline{A}_{ab} - \overline{\delta A}_{ab}) = -\frac{i}{\hbar}\sum_h A_h \delta\widetilde{p}_h(a,b)$ do give an estimation of the uncertainty (randomness) in the distribution $\widetilde{p}_h(a,b)$ of the path contribution. This interpretation hints a map of probability and information theory from phase space of mechanics to Hilbert space of functions. A point in phase space will be mapped on a function in Hilbert space. The paths of real motion in phase space are then mapped one to one on the trajectories of corresponding functions in Hilbert space. The formal probability $\widetilde{p}_h(a,b)$ and varentropy $\widetilde{S}_{ab}$ in Hilbert space are in this way the complex images of the transition probability $p_h(a,b)$ between points and its uncertainty measure, the path entropy $S_{ab}$, in phase space. The physical sense of this complex image of phase space statistics roots in the usual action $A_h(a,b)$ of the mechanics calculated in phase space and used in the least action principle $\overline{\delta A}_{ab} = 0$ of Eq.(17) which implies $\delta(\widetilde{S}_{ab} - \frac{1}{i\hbar}\overline{A}_{ab}) = 0$. As shown in the appendix, if the varentropy takes Shannon form, the Feynman factor follows.

This is the essential concepts of a variational formulation of quantum mechanics which can be included in the DVP framework. However, one should remember that the map "Phase space physics $\Leftrightarrow$ Hilbert space mathematics", or the hypothesis of imaginary diffusion (time or energy) cannot supply the secret of quantum mechanics. It remains a formal tool for axiomatization of quantum theory with least action principle. The true physical reason of this seemingly useful maneuver is still hidden from our knowledge.

## 8) Concluding remarks

From the above examples of derivation of several principles of physics, DVP seems to be the reality behind the apparently different variational principles. The ontological basis of this approach and the dialectical view of the world are not only a long-standing philosophical belief but also a very elementary observed fact of the nature. The only new gradient we introduce is the variational balance. This balance is not the conventional equilibrium usually defined by vanishing difference between the magnitudes of apparently opposing quantities, since the opposite parts considered are not really opposing (conflicting) elements like action/reaction forces which can be forces of same nature. The opposite elements co,sidered in DVP are of incompatible and mutual exclusive nature with contrasting and complementary character such as active/passive, disorder/order, curvature/flatness, apparent/hidden, etc. The use of variational relationship in DVP makes it possible for the balance to be independent of chosen frame of reference. It is expected this axiomatic approach can improve our understanding of physical world and constitutes an alternative way for answering the questions around the origin of variational principles in physics.

One of the attainments of this work is the unification of the axiomatic formulations of classical, statistical and quantum mechanics which have since long time their own fundamental variational principles. Application to electromagnetism theory, mechanical wave theory and geometrical optics should be straightforward since the actions of these cases have been calculated on the basis of the Lagrangian action. However, further effort is needed for its use in relativity theory to calculate Einstein-Hilbert action.

It should also be mentioned that the systems considered in this work are all Hamiltonian systems having well defined Hamiltonian (potential and kinetic energies) at each moment. Consideration of dissipative system and more complex systems would be necessary and useful for further understanding of the approach.

Our final remark is that there are in general many pairs of opposite elements of different nature in a dynamics as mentioned in section 2. One may have impression that, for the DVP description, the choice of the involved pairs of opposite quantities in a motion is crucial. However, from this work, we have seen that the time cumulative energetic definition of $Yi$ and $Ya$, i.e., the time integral of energy, is the most fundamental choice since it yields directly least action principle which gives in turn the virtual work principle and then the maximum entropy principle, although these two principles can be seen independently from other pairs of opposite elements (spatial integral of driving-inertial forces and entropy/expectation). This primordial role of energy integrals in DVP would simplify the application if it was universally confirmed by further investigation.

# *Appenix I*

## *A path probability distribution*

The maximum entropy approach in Eq.(14) does not require any given functional form of $S_{ab}$. Hence different functionals of it are possible[16]. The first possible form which comes to one's mind is the Shannon entropy $S_{ab} = -\sum_h p_h(a,b) \ln p_h(a,b)$. If it is the case, the entropy maximum will yield an exponential path probability distribution $p_h(a,b) \propto e^{-\eta A_h(a,b)}$ [15].

Hence the probability $p_h(a,b)$ for the system to go from a fixed point *a* to a fixed point *b* through a certain path *h* is given by

$$p_h(a,b) = \frac{1}{Z_{ab}} e^{-\eta A_h(a,b)} . \tag{24}$$

where $Z_{ab} = \sum_h e^{-\eta A_h(a,b)}$. This probability distribution has been verified by numerical simulation of diffusion with certain noises[15]. It is also shown that this probability distribution satisfies a Fokker-Planck equation for normal diffusion and is just the transition probability of free Brown motion[15]. Finally the path entropy between two fixed points *a* et *b* can be calculated by [15]

$$S_{ab} = \ln Z_{ab} + \eta \overline{A}_{ab} . \tag{25}$$

This result can be extended to the case where *b* is an unfixed point in the final phase volume *B* of the system at time $t_b$. In this case, the probability $p_h(a,B)$ for the system to go from a fixed point *a* to a certain point *b* (unfixed in the final phase volume *B* of the system at time $t_b$) through a certain path *h* (depending on *a* and *b*) is given by

$$p_h(a,B) = \frac{1}{Z_a} e^{-\eta A_h(a,b)} . \tag{26}$$

where $Z_a = \sum_{b,h} e^{-\eta A_h(a,b)} = \sum_b Z_{ab}$. Hence the path entropy is given by

$$S_{aB} = \ln Z_a + \eta \overline{A}_a . \tag{27}$$

Here $\overline{A}_a = \sum_b \sum_h p_h(a,B) A_h(a,b)$. If we extend still this to the case where neither *a* nor *b* is fixed, we can study the path entropy (the total path uncertainty) between the initial phase volume *A* of the system at time $t_a$ and the final phase volume *B* at time $t_b$. Obviously this is

$$p_h(A,B) = \frac{1}{Z} e^{-\eta A_h(a,b)} . \tag{28}$$

where $Z = \sum_{a,b,h} e^{-\eta A_h(a,b)} = \sum_a Z_a$. The total path entropy between A and B will be

$$S_{AB} = \ln Z \; + \eta \overline{A} . \tag{29}$$

Here $\overline{A} = \sum_a \sum_b \sum_h p_h(A,B) A_h(a,b)$.

If we are interested in the total transition probability $p_{ab}$ between two fixed points *a* and *b*, $p_{ab}$ can be calculated from Eq.(28) by $p_{ab} = \frac{1}{Z} \sum_h e^{-\eta A_h(a,b)} = \frac{Z_{ab}}{Z}$. By using Eq.(25), it can be straightforwardly written

$$p_{ab} = \frac{1}{Z} \exp(S_{ab} - \eta \overline{A}_{ab}) . \tag{30}$$

Using a Legendre transformation $F_{ab} = \overline{A}_{ab} - S_{ab} / \eta = \frac{1}{\eta} \ln Z_{ab}$ which can be called *free action* mimicking the free energy of thermodynamics, we can write $p_{ab} = \frac{1}{Z} \exp(-\eta F_{ab})$.